\documentclass[12pt]{article}

\usepackage{amssymb}
\usepackage{amsmath}

\begin{document}

\title{{\large\bf $\nu - K^0$ Analogy, Dirac-Majorana Neutrino Duality} \\
       {\large\bf and the Neutrino Oscillations}}

\author{{\large E.M. Lipmanov}\\
        {\it 40 Wallingford Rd \#272, Brighton, MA 02135, USA}\\
        {\it E-mail address: \quad elipmanov@yahoo.com}}

\date{}

\maketitle

\begin{abstract}

The intent of this paper is to convey a new primary physical idea of
a Dirac-Majorana neutrino duality in relation to the topical problem
of neutrino oscillations.
In view of the new atmospheric, solar and the LSND neutrino oscillation
data, the Pontecorvo $\nu - K^0$ oscillation analogy is generalized
to the notion of neutrino duality with substantially
different physical meaning ascribed to the long-baseline and the
short-baseline neutrino oscillations.
At the level of CP-invariance, the suggestion of dual neutrino
properties defines the symmetric two-mixing-angle form of
the widely discussed four-neutrino $(2+2)$-mixing scheme, as a result
of the lepton charge conservation selection rule and a minimum of two
Dirac neutrino fields. With neutrino duality, the two-doublet
structure of the Majorana neutrino mass spectrum is a vestige of
the two-Dirac-neutrino origin. The fine neutrino mass doublet
structure is natural because it is produced by a lepton charge symmetry
violating perturbation on a zero-approximation system of two twofold
mass-degenerate Dirac neutrino-antineutrino pairs. A set of inferences
related to the neutrino oscillation phenomenology in vacuum is considered.

\end{abstract}

\section{Introduction}
\label{sec:introduction}

In view of the neutrino oscillation data, a simple $(2+2)$-form of the
four-neutrino mixing phenomenology~\cite{Bilenky-99}
is considered in~\cite{Lipmanov-98} which is restricted to CP-conservation
and based on an, initiated by Pontecorvo~\cite{Pontecorvo}, extended
neutrino analogy with the duality behavior of the hadronic system of the
neutral kaons~\cite{Lipmanov-99}. Unlike the kaons, the neutrinos are
elementary particles, leptons, and this lepton-hadron oscillation analogy
prompted the suggestion of a new physical notion, the Dirac-Majorana
neutrino duality. As it is known, a carrying lepton charge four-component
Dirac field can be represented as a maximal mixing superposition of two
two-component Majorana fields with equal masses and opposite CP-parities,
e.g.~\cite{Bilenky-99}. ``Duality'' means that massive neutrinos behave
either as Dirac particles, or antiparticles, or as pairs of Majorana ones
depending on the specific physical
phenomena in which they are observed, and therefore the Majorana neutrinos
come only in pairs with opposite CP-parities.

Beyond the formal differences, the suggested Dirac-Majorana
duality properties of the neutrinos and the known duality
properties of the neutral kaons have certain similar roots in
particle physics. A massive charged field, ($K^0$-meson with
strangeness, or Dirac neutrino fields with lepton charge) is a
maximal mixing superposition of two ``neutral'' fields with equal
masses, the unification is protected by symmetry. The
``long-baseline'' particle oscillations (oscillations of
strangeness, or lepton charge) afford a universal mechanism for
the transformation of very small perturbation effects of a real
short-range charge symmetry violating interaction into certain
large observable physical effects. It is because the charge
symmetry violation renders the mentioned above charged field
superposition phase-unlocked what leads via the long-baseline
oscillations to large observable effects at large distances~$L$
from the particle production vertex with a relativistic
dimensional relation $L \sim E / \Delta m^2$ where $\Delta m^2$ is
the mass-squared difference of the emerging pair of real neutral
fields ($K^0_1$, $K^0_2$-mesons, or a pair of Majorana neutrino
fields) and~$E$ is their energy. Hence the origin of the suggested
Dirac-Majorana neutrino duality can be analogous to the known
origin of the neutral kaon duality.

Just because particle-antiparticle oscillation phenomena are well known
in physics, they cannot be easily dismissed in
the neutrino oscillation phenomenology. With such oscillations, the
necessary condition that the charge of a particle maintains its explicit
physical meaning reads $\Delta m^2 \ll m^2$, i.e. the charge violating
interaction must be a small perturbation effect (particle oscillations
with $\Delta m^2 \gtrsim m^2$,
if possible, should be charge conserving ``short-baseline'' oscillations,
what is a different kind of particle oscillations). The long-baseline
oscillations generate maximal charge violating effects no matter how weak
are the new charge violating interactions. If the latter vanish, the
long-baseline oscillation length would just grow infinitely, with the
relevant particles being definitely Dirac fermions. Otherwise, they have
dual properties. The neutrinos can be described as
Dirac neutrino or antineutrino states in the semiweak interactions
while being described as pairs of Majorana neutrino states in the
long-baseline neutrino oscillations. No consistent theory of such a
fundamental Dirac-Majorana neutrino duality is known as yet,
also~\cite{Ellis-PLB99}.
But the neutrino oscillation data seem to point out new physics, and this
notion is not excluded as yet. The notion of Dirac-Majorana neutrino duality
at the approximation level of CP-invariance leads below to a physically
motivated minimal neutrino mixing ansatz (the simplest symmetric form of the
so-called $(2+2)$-scheme) with not more than two mixing angles, it conforms
naturally to a number of well known neutrino oscillation data. Indications
that may be in apparent disagreement with the minimal model of Dirac-Majorana
neutrino duality, Sec.~\ref{sec:DM-ansatz}, come from the recent
Super-Kamiokande solar neutrino oscillation data, Sec.~\ref{sec:fitting}.

\section{Dirac-Majorana neutrino duality ansatz}
\label{sec:DM-ansatz}

The three weak interaction neutrino eigenstates $\nu_e$, $\nu_\mu$
and~$\nu_\tau$ in the known charged and neutral weak interaction lepton
currents of the SM, plus one sterile neutrino~$\nu_s$, and also the weak
interaction eigenstates of the corresponding antineutrinos
$\tilde \nu_e \equiv (\nu_e)^c$, $\tilde \nu_\mu \equiv (\nu_\mu)^c$,
$\tilde \nu_\tau \equiv (\nu_\tau)^c$, and $\tilde \nu_s \equiv (\nu_s)^c$,
are represented by the following superposition ansatz~\cite{Lipmanov-99}:
\begin{alignat}{2}
\nu_e &= \left [ \nu^{\rm D}_{\rm I} \cos \vartheta
          + \nu^{\rm D}_{\rm II} \sin \vartheta \right ]_{\rm L},
& \qquad
\tilde \nu_e &= \left [ (\nu^{\rm D}_{\rm I})^c \cos \vartheta
               + (\nu^{\rm D}_{\rm II})^c \sin \vartheta \right ]_{\rm R},
\label{eq:nu_e} \\
\nu_\mu &= \left [ - \nu^{\rm D}_{\rm I} \sin \vartheta
            + \nu^{\rm D}_{\rm II} \cos \vartheta \right ]_{\rm L},
& \qquad
\tilde \nu_\mu &= \left [ - (\nu^{\rm D}_{\rm I})^c \sin \vartheta
                 + (\nu^{\rm D}_{\rm II})^c \cos \vartheta \right ]_{\rm R},
\label{eq:nu_mu} \\
\nu_\tau &= \left [ (\nu^{\rm D}_{\rm I})^c \sin \varphi
             + (\nu^{\rm D}_{\rm II})^c \cos \varphi \right ]_{\rm L} ,
& \qquad
\tilde \nu_\tau &= \left [ \nu^{\rm D}_{\rm I} \sin \varphi
                  + \nu^{\rm D}_{\rm II} \cos \varphi \right ]_{\rm R} ,
\label{eq:nu_tau} \\
\nu_s &= \left [ (\nu^{\rm D}_{\rm I})^c \cos \varphi
          - (\nu^{\rm D}_{\rm II})^c \sin \varphi \right ]_{\rm L} ,
& \qquad
\tilde \nu_s &= \left [ \nu^{\rm D}_{\rm I} \cos \varphi
               - \nu^{\rm D}_{\rm II} \sin \varphi \right ]_{\rm R} ,
\label{eq:nu_s}
\end{alignat}
Here, $\nu^{\rm D}_{\rm I}$ and $\nu^{\rm D}_{\rm II}$ are two
carrying lepton charge four-component Dirac neutrino fields with masses
$m^{\rm D}_{\rm I}$ and $m^{\rm D}_{\rm II}$ and with the mass
Lagrangian term
\begin{equation}
\Delta L^{\rm D} =
- \frac{1}{2} \, \sum_i m^{\rm D}_i \left [ \,
\overline{\nu^{\rm D}_i} \nu^{\rm D}_i  +
\overline{(\nu^{\rm D}_i)^c} (\nu^{\rm D}_i)^c \,
\right ] ,
\qquad i = {\rm I}, {\rm II}.
\label{eq:mass-termD}
\end{equation}
The symbols $(\nu^{\rm D}_i)^c$ denote the Dirac antineutrino
fields and the superscript~$c$ means charge conjugation. The~L and~R
subscripts denote left and right chirality states of the neutrinos in
the weak interaction currents, which are defined by the projection
operators $(1 - \gamma_5)/2$ and $(1 + \gamma_5)/2$, respectively.

With the ansatz in Eqs.~(\ref{eq:nu_e})--(\ref{eq:nu_s}), the
lepton charge is conserved in the semiweak interactions of the SM
with the definition of the seven ``leptons'' ($l_k = + 1$, $k = 1
\div 7$):
\begin{equation}
\nu_e, \, \, \nu_\mu, \, \,  \tilde \nu_\tau, \, \, \tilde \nu_s, \, \,
e^-, \, \, \mu^-, \, \, \tau^+,
\label{eq:leptons-7}
\end{equation}
and the seven ``antileptons'' ($l_k = -1$, $k = 1 \div 7$):
\begin{gather}
\tilde \nu_e,  \, \, \tilde \nu_\mu,  \, \, \nu_\tau,  \, \,
\nu_s,  \, \, e^+,  \, \, \mu^+,  \, \, \tau^-.
\tag{\ref{eq:leptons-7}$'$}
\end{gather}

The selection rule of lepton charge conservation in the semiweak
interactions with the lepton charge definition in Eq.~(\ref{eq:leptons-7})
leads to the important no-mixing of the Dirac neutrino mass eigenstates
$\nu^{\rm D}_i$ with the Dirac antineutrino mass eigenstates
$(\nu^{\rm D}_i)^c$ in the ansatz~(\ref{eq:nu_e})--(\ref{eq:nu_s})
and therefore to the physically motivated restriction to only two mixing
angles~$\vartheta$ and~$\varphi$.

In contrast to the conservation of the separate electron,
muon and tauon numbers, the lepton charge conservation with the
definition~(\ref{eq:leptons-7}) is not violated
by the Dirac neutrino masses, Eq.~(\ref{eq:mass-termD}), in the electroweak
perturbation theory.

The definition of the lepton charge in Eq.~(\ref{eq:leptons-7}) is
rather uncommon. Usually all three lepton flavor pairs ($\nu_e$,
$e$), ($\nu_\mu$, $\mu$) and ($\nu_\tau$, $\tau$) are implied to
have the same lepton charge, e.g. $l = +1$. The suggestion that
the three neutrinos are represented (beyond the SM) by three
independent four-component Dirac fields would lead then to three
sterile neutrinos as the right components of the Dirac neutrino
fields with zero weak isospin and hypercharge. With the definition
of the lepton charge~(\ref{eq:leptons-7}) and the
ansatz~(\ref{eq:nu_e})--(\ref{eq:nu_s}) and~(\ref{eq:mass-termD}),
we introduced not three but only two four-component Dirac
neutrinos and so there is here only one sterile neutrino. The new
$SU_L (2) \times U (1)$ electroweak symmetry representation and
lepton charge content of the neutrinos in the
ansatz~(\ref{eq:nu_e})--(\ref{eq:nu_s}) get more transparent in
the limit of equal mixing angles $\varphi = - \vartheta$. In this
case we can define two four-component Dirac neutrinos~$\nu^e$
and~$\nu^\mu$ in a one-parameter economical version of the
ansatz~(\ref{eq:nu_e})--(\ref{eq:nu_s}),
\begin{eqnarray}
&&
\nu_e = \nu^e_{\rm L}
\,
(\tilde \nu_e = \tilde \nu^e_{\rm R}),
\qquad
\tilde \nu_s = \nu^e_{\rm R}
\,
(\nu_s = \tilde \nu^e_{\rm L}),
\nonumber \\
&&
\nu_\mu = \nu^\mu_{\rm L}
\,
(\tilde \nu_\mu = \tilde \nu^\mu_{\rm R}),
\qquad
\tilde \nu_\tau = \nu^\mu_{\rm R}
\,
(\nu_\tau = \tilde \nu^\mu_{\rm L}),
\label{eq:neutrinos} \\
&&
\nu^e \equiv \nu^{\rm D}_{\rm I} \cos \vartheta
           + \nu^{\rm D}_{\rm II} \sin \vartheta ,
\qquad
\nu^\mu \equiv - \nu^{\rm D}_{\rm I} \sin \vartheta
               + \nu^{\rm D}_{\rm II} \cos \vartheta .
\nonumber
\end{eqnarray}
Both the left and right components of the four-component Dirac
``muon'' neutrino~$\nu^\mu$ are active in the weak interactions
and represent the up- and down-components of the $SU_L (2)$-doublets
$(\nu^\mu, \mu^-)_{\rm L}$ and $(\tau^+, \nu^\mu)_{\rm R}$, respectively.
On the other hand, only the left component of the four-component Dirac
``electron'' neutrino~$\nu^e$ in the electroweak doublet
$(\nu^e, e^-)_{\rm L}$ is active in the weak interactions, whereas its
right component $\nu^e_{\rm R}$ remains a sterile neutrino in the SM.
In contrast to the case of massless neutrinos in the SM, the Dirac mass
Lagrangian term~(\ref{eq:mass-termD}) generates transitions
$\nu^\mu_{\rm L} \leftrightarrow \nu^\mu_{\rm R}$
and $\nu^e_{\rm L} \leftrightarrow \nu^e_{\rm R}$, i.e. it makes possible
the transitions of the muon neutrino into the tau antineutrino,
$\nu_\mu \leftrightarrow \tilde \nu_\tau$, and also transitions of the
electron neutrino into sterile antineutrino, which will be realized by
the involvement of the Dirac masses in the weak interactions.
In the more general mixing with $\vartheta \neq - \varphi$, as in
Eqs.~(\ref{eq:nu_e})--(\ref{eq:nu_s}), the Dirac neutrino masses generate
all four possible types of the lepton charge conserving transitions
in the weak interactions
$(\nu_e, \nu_\mu) \leftrightarrow (\tilde \nu_\tau, \tilde \nu_s)$.
With the definition in Eq.~(\ref{eq:leptons-7}), interesting though
extremely suppressed (factor $m_\nu^2 /E_\nu^2$) lepton charge conserving
reactions, e.g.
\begin{displaymath}
\nu_\mu + (A, Z) \to (A, Z - 1) + \tau^+ ,
\end{displaymath}
are possible through the involvement of the Dirac neutrino masses in the
lepton weak interactions. It is appropriate to note here that massive
neutrinos are produced in the weak interactions with not complete
longitudinal polarization. In our neutrino mixing
pattern~(\ref{eq:nu_e})--(\ref{eq:nu_s}), it means that the
incoming~$\nu_\mu$ in the reaction above is an effective state including
the muon neutrino plus a very small admixture of the tau and sterile
antineutrinos. Because of the suppression by the very small neutrino masses,
the violation of the $(e - \mu) - \tau$ universality in the definition of
the lepton charge in Eq.~(\ref{eq:leptons-7}) does not disagree with any
known data.

The data of the atmospheric, solar and the LSND experiments can be
explained with the following values of the neutrino mass-squared
differences~\cite{Bilenky-99}
\begin{eqnarray}
\Delta m_1^2 & = & m_1^{'2} - m_1^2 = \Delta m^2_{\rm solar}
\sim 10^{-10}(\mbox{Vac}), \, \, \mbox{or}
\sim 10^{-5}  \, (\mbox{MSW}) \,  \, \mbox{eV}^2,
\nonumber \\
\Delta m_2^2 & = & m_2^{'2} - m_2^2 = \Delta m^2_{\rm atm}
\sim 10^{-3} \div 10^{-2}  \, \mbox{eV}^2,
\label{eq:mass-differences} \\
\Delta m_{12}^2 & \cong & m_2^2 - m_1^2
\sim 1  \, \mbox{eV}^2
\nonumber
\end{eqnarray}
with a possible scheme of the neutrino mass spectrum
\renewcommand{\theequation}{\arabic{equation}A}
\begin{equation}
\overset{\rm solar}{\overbrace{m_1 < {m'}_1}} \ll
\overset{\rm atm}{\overbrace{m_2 < {m'}_2}}.
\label{eq:mass-spectrum}
\end{equation}
\renewcommand{\theequation}{\arabic{equation}}
and another possible scheme~(9B) where the positions of the
``solar'' and the ``atm'' doublet splittings of Eq.~(\ref{eq:mass-spectrum})
are interchanged.

To describe the neutrino mass data of Eq.~(\ref{eq:mass-differences})
in accordance with the single phenomenological approach of the
$\nu - K^0$ analogy, we should introduce an effective mass Lagrangian
term~$\Delta L^{\rm M}$, called Majorana term, with maximal lepton charge
violation $\Delta l = \pm 2$,
\begin{equation}
\Delta L^{\rm M} = \frac{1}{2} \sum_i m_i^{\rm M}
\left [ \, \overline{\nu^{\rm D}_i} (\nu^{\rm D}_i)^c
+ \overline{(\nu^{\rm D}_i)^c} \nu^{\rm D}_i \right ] ,
\qquad
0 < m_i^{\rm M} \ll m_i^{\rm D},
\qquad i = {\rm I}, \, {\rm II}.
\label{eq:mass-termM}
\end{equation}
It has the physical meaning of a very small perturbation of the twofold
mass-degenerate Dirac neutrino-antineutrino zero order approximation system,
Eq.~(\ref{eq:mass-termD}). This type of exact neutrino mass-degeneracy
is certainly a natural zero-order approximation because it is protected
by the CPT-invariance. To consider the lepton charge violating interaction
as a small perturbation effect is necessary for modelling the Dirac-Majorana
neutrino duality, see Sec.~\ref{sec:introduction}.

The simplest admissible texture of the neutrino interactions, used here,
reads: the sole location of lepton charge violation in the lepton
interactions is in the left-right symmetric effective Majorana mass
term~(\ref{eq:mass-termM}) with physical meaning of a small perturbation,
whereas the sole location of the left-right asymmetry is in the lepton
charge conserving weak interactions. The left-right symmetry of the Majorana
term~(\ref{eq:mass-termM}) leads below to the maximal mixing of the Majorana
neutrinos. The perturbative status of the Majorana term~(\ref{eq:mass-termM})
leads to the fine doublet structure of the neutrino mass spectrum.

It should be noted that the Dirac and Majorana neutrino mass Lagrangian
terms~(\ref{eq:mass-termD}) and~(\ref{eq:mass-termM}) are not compatible
with the minimal Higgs mechanism of the SM electroweak theory. They are
introduced here as a continuation of the neutrino mixing
ansatz~(\ref{eq:nu_e})--(\ref{eq:nu_s}) in an attempt to scheme the new
physics beyond the SM, which could be the source of the suggested dual
Dirac-Majorana neutrino properties; the Majorana mass
term~(\ref{eq:mass-termM}) being analogous to the ($K^0_1$, $K^0_2$) mass
Lagrangian term which schemes the relevant perturbative
$K_0 \leftrightarrow \bar K_0$ effect of the weak interactions. As follows,
the neutrino mixing ansatz in~(\ref{eq:nu_e})--(\ref{eq:nu_s}),
(\ref{eq:mass-termD}) and~(\ref{eq:mass-termM}) leads uniquely to the
symmetric form~\cite{Lipmanov-98} of the widely discussed,
e.g.~\cite{(2+2)-mixing scheme}, four-neutrino $(2 + 2)$-mixing scheme.

The total neutrino mass Lagrangian term can be rewritten in another form
\begin{eqnarray}
\Delta L^\nu & = & \Delta L^{\rm D} + \Delta L^{\rm M}
= - \frac{1}{2} \sum_k (m_k \, \bar \nu_k \nu_k +
m^\prime_k \, \bar \nu_k^\prime \nu_k^\prime ), \qquad k = 1, 2,
\label{eq:Lagr-total} \\
m_k & = & (m^{\rm D}_k - m^{\rm M}_k ), \, \, \,
m^\prime_k = (m^{\rm D}_k + m^{\rm M}_k), \, \, \,
m^\prime_k - m_k \ll m_k .
\nonumber
\end{eqnarray}
Here, ($\nu_k$, $\nu^\prime_k$), $k = 1, 2$, are two pairs of truly
neutral Majorana neutrino mass eigenstates (with CP-invariance they
are stationary states). The Majorana neutrino states are
related to the primary neutrino and antineutrino Dirac mass eigenstates
$\nu^{\rm D}_{\rm I}$, $\nu^{\rm D}_{\rm II}$ and $(\nu^{\rm D}_{\rm I})^c$,
$(\nu^{\rm D}_{\rm II})^c$:
\begin{equation}
\nu_k = \frac{1}{\sqrt 2}
\left [ \nu^{\rm D}_k + (\nu^{\rm D}_k)^c \right ] ,
\qquad
\nu^\prime_k = \frac{1}{\sqrt 2}
\left [ \nu^{\rm D}_k - (\nu^{\rm D}_k)^c \right ] .
\label{eq:nu-eigenstate1}
\end{equation}
The CP-parities of these Majorana neutrinos are opposite in each of the
two doublets, see also~\cite{CP-parity}.

After the small perturbation by the Majorana mass term~(\ref{eq:mass-termM})
is switched on, the Majorana neutrino states~$\nu_k$ and~$\nu^\prime_k$
are the new stationary states, while~$\nu^{\rm D}_k$ and~$(\nu^{\rm D}_k)^c$
in Eq.~(\ref{eq:nu-eigenstate1}) are the new nonstationary neutrino states,
which are not Dirac states with lepton charge any more, comp.
Sec.~\ref{sec:introduction}. It is because of phase-unlocking effect
due to the different masses of the Majorana neutrinos~$\nu_k$
and~$\nu^\prime_k$ in the superpositions
\begin{gather}
\nu_k^s = \frac{1}{\sqrt 2} (\nu_k + \nu^\prime_k) ,
\qquad
\nu_k^a = \frac{1}{\sqrt 2} (\nu_k - \nu^\prime_k) ,
\qquad
k = 1, 2.
\tag{\ref{eq:nu-eigenstate1}$^\prime$}
\end{gather}
The superscripts~$s$ and~$a$ here denote symmetric and antisymmetric
states under the interchange of the Majorana neutrinos~$\nu_k$
and~$\nu^\prime_k$~\cite{Lipmanov-98}. At distances~$L$ from the
neutrino production vertex in the semiweak interactions, which are
much shorter than the long-baseline oscillation length, $\nu^s_k$
and~$\nu^a_k$ remain nearly Dirac neutrino and antineutrino states.
It is mainly because of two factors: one is the maximal mixing of
the Majorana neutrinos~$\nu_k$ and~$\nu^\prime_k$
in~(\ref{eq:nu-eigenstate1}$^\prime$) as a vestige
of their primary Dirac neutrino origin in the semiweak interactions,
and the other is negligible physical effects of lepton charge
violation owing to the small differences of the Majorana neutrino
phases in the neutrino mass doublets at the stated distances~$L$.
The Majorana neutrino states in Eq.~(\ref{eq:nu-eigenstate1}) correspond
to the truly neutral meson states~$K^0_1$ and~$K^0_2$ in the used here
analogy, while the Dirac neutrinos~$\nu^{\rm D}_i$ and
antineutrinos~$\big ( \nu^{\rm D}_i \big )^c$ correspond to the
primary neutral kaons which are produced in the strong interactions with
definite strangeness as~$K^0$ or~$\bar K^0$ mesons and become the symmetric
and antisymmetric superpositions, i.e. $K^0 = (K^0_1 + K^0_2)/\sqrt 2$
and $\bar K^0 = (K^0_1 - K^0_2)/\sqrt 2$, at finite distances~$L$ from
the production vertex, comp. Eq.~(\ref{eq:nu-eigenstate1}$^\prime$).

Unlike the formal description in the review~\cite{Bilenky-99}, where
the charge parities of the Majorana neutrinos are always chosen to be
$C = + 1$, the choice here of opposite charge parities $C = + 1$ and
$C = - 1$ for the Majorana neutrinos~$\nu_k$ and~$\nu^\prime_k$ respectively
proved to be more convenient for the purpose of modelling the Dirac-Majorana
neutrino duality.

Lepton charge violating reactions, e.g. neutrinoless double $\beta$-decay,
are possible because of the Majorana neutrino mass doublet
splittings, they are naturally suppressed by factors much smaller than in
the mentioned above possible lepton charge conserving
$\nu_\mu \leftrightarrow \tilde \nu_\tau$ transformations because of
the relation $\Delta m_1^2 \ll m_1^2$.

As a rule, the doublet structure of a quantum system reveals an
underlying broken symmetry. There is no exception here: the
doublet structure of the physical Majorana neutrino mass spectrum
reveals the broken, by the perturbative effective Majorana term
$\Delta L^{\rm M}$, lepton charge symmetry of the semiweak
interactions. The stabilizing condition for the fine two-doublet
structure of the Majorana neutrino mass spectrum here is the
protected by CPT-invariance mass-degeneracy of the
zero-order approximation Dirac neutrino-antineutrino system.
This fine structure of the Majorana neutrino mass spectrum is natural:
if~$\Delta m_1^2$ and~$\Delta m_2^2$ were zero, the lepton charge
symmetry would be restored.

The formal definition of the nearly Dirac neutrino~$\nu^s_k$ and
antineutrino~$\nu^a_k$ fields is close to the definition of the
``pseudo-Dirac neutrinos'' in Ref.~\cite{Wolfenstein-81}.
The pseudo-Dirac neutrino field approximation is analogous
to the well known effective approximation of carrying strangeness~$K^0$
and~$\bar K^0$ meson states of the strong interactions so far as we count
the weak interactions as very small perturbations,
$m (K^0_2) - m (K^0_1) \ll m (K^0)$. Just as the~$K^0$ and the~$\bar K^0$
meson states are produced in the strong interactions with definite
strangeness (charge) whereas the dual truly neutral meson states~$K^0_1$
and~$K^0_2$ show up in the $K^0$-oscillations, the~$\nu^{\rm D}_i$
and~$(\nu^{\rm D}_i)^c$ neutrino states are produced in the standard weak
interactions with definite lepton charge as Dirac neutrinos whereas the
dual truly neutral Majorana neutrino states~$\nu_k$ and~$\nu^\prime_k$
show up in the long-baseline neutrino oscillations.
The significant difference is that with
three lepton flavors we need two massive Dirac neutrino
fields~$\nu^{\rm D}_{\rm I}$ and~$\nu^{\rm D}_{\rm II}$
(and Dirac antineutrino fields~$(\nu^{\rm D}_{\rm I})^c$
and~$(\nu^{\rm D}_{\rm II})^c$) and
unlike the $K^0$-case there is a possibility of their mixing.
Neutrino mixing in Eqs.~(\ref{eq:nu_e})--(\ref{eq:nu_s}), with different
mixing angles~$\vartheta$ and~$\varphi$ for the two chiral neutrino
states~$\nu^{\rm D}_{i {\rm L}}$ [$(\nu^{\rm D}_i)^c_{\rm R}$],
$i = {\rm I}, {\rm II}$, and the two chiral antineutrino states
$(\nu^{\rm D}_i)^c_{\rm L}$ [$\nu^{\rm D}_{i {\rm R}}]$,
$i = {\rm I}, {\rm II}$, respectively, is considered below. The condition
of different mixings for the left and right components of the two
four-component neutrinos~$\nu^{\rm D}_{\rm I}$ and~$\nu^{\rm D}_{\rm II}$
is here a part of the left-right asymmetry of the lepton weak interactions.
In the one-parameter version~(\ref{eq:neutrinos}) of the present model with
the relation $\varphi = - \vartheta$ this additional left-right asymmetry
disappears.

The sterile neutrino occurrence is here a necessary result of two
conditions: 1) there is the lepton charge conservation in the
semiweak interactions with three lepton flavors and 2) the
neutrinos represented by the minimum of primary two massive
four-component Dirac fields. Because of these conditions and the
maximal parity nonconservation in the neutrino weak interactions,
there must be one two-component nonactive neutrino degree of
freedom, which must be imparted as a sterile neutrino in both the
short-baseline and the long-baseline neutrino oscillations, see
equations below.

The long-baseline neutrino oscillations, including the atmospheric
and solar ones, are effective four-component Dirac
neutrino-antineutrino lepton charge oscillations (i.e. left Dirac
neutrino $\leftrightarrow$ left Dirac antineutrino, or right Dirac
neutrino $\leftrightarrow$ right Dirac antineutrino,
e.g.~(\ref{eq:neutrinos})), as a result of neutrino duality;
beyond the formal differences, these lepton charge oscillations
have the same physical meaning as the oscillations of strangeness
in the $K^0$-meson oscillations: $K^0 \leftrightarrow (K^0_1 +
K^0_2) / \sqrt 2, \, \Delta m \neq 0 \leftrightarrow \, <{\rm
oscillations}, \, \Delta t \neq 0> \, \leftrightarrow (K^0_1 -
K^0_2) / \sqrt 2 \leftrightarrow \bar K^0$, as a result of neutral
kaon duality. On the other hand, the short-baseline neutrino
oscillations are mainly lepton charge conserving effective
two-Dirac-neutrino oscillations resulting in the transformations
$\nu_\mu \leftrightarrow \nu_e$ with the oscillation amplitude
$\sin^2 2\vartheta$, and $\nu_\tau \leftrightarrow \nu_s$ with the
oscillation amplitude $\sin^2 2\varphi$. These short-baseline
oscillations have no $K^0$-analogy because of the absence of a
relevant second $K$-generation, they would remain mainly unchanged
even if the neutrino mass matrix were of the regular Dirac type
with no neutrino mass doublet splittings. The main inference here
is that only the two mentioned types of the short-baseline
neutrino oscillations are possible: because of lepton charge
conservation in the oscillations of this type, there are no
short-baseline $\nu_\mu, \nu_e \leftrightarrow \nu_\tau, \nu_s$,
oscillation transformations. This inference is supported by the
positive LSND indications~\cite{Athanassopoulos} and also by the
negative data of the CHORUS and NOMAD experiments~\cite{CHORUS}.

\section{Fitting to the neutrino oscillation data}
\label{sec:fitting}

The equations for the neutrino oscillation probabilities below resulted
from Eqs.~(\ref{eq:nu_e}), (\ref{eq:nu_mu}), (\ref{eq:nu_tau})
and~(\ref{eq:nu_s}) after revealing there (and also in
Eq.~(\ref{eq:neutrinos})) the Dirac-Majorana neutrino duality condition:
\begin{equation}
\nu^{\rm D}_i \to \nu^s_i = \frac{1}{\sqrt 2}
\left ( \nu_i + \nu^\prime_i \right ),
\qquad
(\nu^{\rm D}_i)^c \to \nu^a_i = \frac{1}{\sqrt 2}
\left ( \nu_i - \nu^\prime_i \right ),
\label{eq:substitutions}
\end{equation}
in accordance with Eq.~(\ref{eq:nu-eigenstate1}$^\prime$). The probabilities
are written in terms of the neutrino mass scheme~(\ref{eq:mass-spectrum}),
but the conclusions do not depend on this choice.

The probability of the $\nu_\mu \leftrightarrow \nu_e$ oscillations reads
\begin{equation}
W(\nu_\mu \leftrightarrow \nu_e) = \sin^2 2\vartheta
\left [ < \sin^2 (\delta m_{12}^2) > - \frac{1}{4} \sin^2 (\delta m_1^2)
- \frac{1}{4} \sin^2 (\delta m_2^2) \right ],
\label{eq:Wmu-e}
\end{equation}
where the symbol $< \quad >$ in the first term denotes the arithmetic mean
value of the appropriate four terms related to the four ``large''
mass-squared differences among the two different neutrino mass doublets.
The notations in Eq.~(\ref{eq:Wmu-e}) and below are
\begin{equation}
(\delta m_{ij}^2) \equiv \frac{\Delta m_{ij}^2 \, L}{4 E} , \quad
\Delta m_{ij}^2 \equiv m_j^2 - m_i^2, \quad
(\delta m_1^2) \equiv (\delta m_{1 1^\prime}^2), \quad
(\delta m_2^2) \equiv (\delta m_{2 2^\prime}^2),
\label{eq:mass-notations}
\end{equation}
with $i, j = 1, 1^\prime, 2, 2^\prime$. $E$ is the neutrino energy
and~$L$ is the distance from the neutrino source to the detector.
The probabilities of the $\nu_\mu \leftrightarrow \nu_e$ transformations
are determined by one factor $\sin^2 2\vartheta$, which is the effective
LSND ``two-neutrino'' oscillation amplitude with an
estimation~\cite{Athanassopoulos}
\begin{equation}
\sin^2 2\vartheta \approx 2 \times 10^{-3} \div 4 \times 10^{-2} .
\label{eq:sin-2theta}
\end{equation}
The second and third terms in Eq.~(\ref{eq:Wmu-e}) give small contributions
to the total~$\nu_\mu$- and~$\nu_e$-disappearance atmospheric and solar
oscillation probabilities (comp. Eqs.~(\ref{eq:Wmu-tau'st'e})
and~(\ref{eq:We-tau'st'mu})) from the channels~$\nu_\mu \to \nu_e$
and~$\nu_e \to \nu_\mu$, respectively. These negative sign terms in the
probability $W(\nu_\mu \leftrightarrow \nu_e)$ indicate subtraction of
the long-baseline lepton charge disappearance probabilities, what is a
necessary result of the lepton charge conservation in the short-baseline
oscillations in the present model.

The probability of the short-baseline $\nu_\tau \leftrightarrow \nu_s$
oscillation transformations follows from Eq.~(\ref{eq:Wmu-e}) after the
substitution $\vartheta \to \varphi$,
\begin{equation}
W(\nu_\tau \leftrightarrow \nu_s) =
\frac{\sin^2 2\varphi}{\sin^2 2\theta} \,
W(\nu_\mu \leftrightarrow \nu_e).
\label{eq:Wtau-st}
\end{equation}
A direct estimation of the mixing angle~$\varphi$ in Eq.~(\ref{eq:Wtau-st})
could be by the measurement of the $\nu_\tau$-survival probability
in the short-baseline tau-neutrino oscillation experiment. It is not
available at present, but the recent detection of the tau-neutrino at
Fermilab~\cite{nu_tau detection} should eventually make possible such an
experiment.

The probability of the $\nu_\mu \leftrightarrow \nu_\tau$ oscillations reads
\begin{eqnarray}
W(\nu_\mu \leftrightarrow \nu_\tau) & = &
\cos^2 \vartheta \cos^2 \varphi \sin^2 (\delta m_2^2) +
\sin^2 \vartheta \sin^2 \varphi \sin^2 (\delta m_1^2)
\label{eq:Wmu-tau} \\
& + & \frac{1}{4} \sin 2 \vartheta \sin 2 \varphi
\left [
\sin^2 (\delta m_{1^\prime 2^\prime}^2) + \sin^2 (\delta m_{12}^2)
- \sin^2 (\delta m_{1 2^\prime}^2) - \sin^2 (\delta m_{1^\prime 2}^2)
\right ].
\nonumber
\end{eqnarray}

The probability of the $\nu_\mu \leftrightarrow \nu_s$ oscillations
can be obtained from Eq.~(\ref{eq:Wmu-tau}) by the substitution
$\varphi \to \varphi + \pi / 2$, and so the probability of the
$\nu_\mu \to \nu_\tau + \nu_s$ transformations,
\begin{equation}
W(\nu_\mu \to \nu_\tau + \nu_s) = \cos^2 \vartheta \sin^2 (\delta m_2^2)
+ \sin^2 \vartheta \sin^2 (\delta m_1^2),
\label{eq:Wmu-tau'st}
\end{equation}
is independent of the second mixing angle~$\varphi$ and of
the neutrino mass doublet separation~$\Delta m_{12}$.
Eq.~(\ref{eq:Wmu-tau'st}) describes the probability of the
long-baseline $\nu_\mu$ oscillation transformations.
With the contribution from Eq.~(\ref{eq:Wmu-e}), the total
probability of the $\nu_\mu$ oscillation transformations reads
\begin{equation}
W(\nu_\mu \to \nu_\tau + \nu_s + \nu_e) =
\cos^4 \vartheta \sin^2 (\delta m_2^2) +
\sin^4 \vartheta \sin^2 (\delta m_1^2) +
\sin^2 2 \vartheta < \sin^2 (\delta m_{12}^2) >.
\label{eq:Wmu-tau'st'e}
\end{equation}
The first term in Eq.~(\ref{eq:Wmu-tau'st'e}) is the dominant one,
it gives the main part of the muon neutrino disappearance probability
in the atmospheric long-baseline $\nu_\mu$-neutrino oscillations with
the large amplitude
\begin{equation}
A_{\rm atm} \cong \cos^4 \vartheta \cong 1.
\label{eq:A-atm}
\end{equation}
It comes mainly from the transformations~$\nu_\mu \to \nu_\tau$
and~$\nu_\mu \to \nu_s$, their partial contributions carry the
factors~$\cos^2 \varphi$ and~$\sin^2 \varphi$, respectively.
At $\varphi = 45^\circ$ the ratio of these contributions
$\cot^2 \varphi$ is near to one. The contribution from the
transformations $\nu_\mu \to \nu_e$ in Eq.~(\ref{eq:Wmu-tau'st'e})
is a small short-baseline effect. The recent data indications
in favor of the $\nu_\mu \to \nu_\tau$ transitions in the atmospheric
$\nu_\mu$-oscillations~\cite{Fukuda} seem to exclude a dominant
contribution from the $\nu_\mu \to \nu_s$ mode in these oscillations.

The probability of the $\nu_\mu$-survival oscillations and the
$\nu_\mu \to \nu_\tau$ appearance transformations in the coming
long-baseline accelerator $\nu_\mu$-oscillation experiments should be
\begin{equation}
W (\nu_\mu \to \nu_\mu) \cong
\cos^2 \frac{\Delta m_2^2 \, L}{4 E},
\qquad
W (\nu_\mu \to \nu_\tau) \cong \cos^2 \varphi \,
\sin^2 \frac{\Delta m_2^2 \, L}{4 E},
\label{eq:Wmu-mu}
\end{equation}
where~$L$ is the distance from the neutrino source to the detector.
The results of these experiments should lead to a clear estimation
of the mixing angle~$\varphi$ through the term $\cos^2 \varphi$
in Eq.~(\ref{eq:Wmu-mu}).

The probability of the~$\nu_\tau$ oscillation transformations
$\nu_\tau \to \nu_\mu + \nu_e + \nu_s$ follows from
Eq.~(\ref{eq:Wmu-tau'st'e}) after the substitution $\vartheta \to \varphi$.

The probability of the $\nu_e \to \nu_s$ and $\nu_e \to \nu_s + \nu_\tau$
long-baseline oscillation transformations can be obtained from
Eqs.~(\ref{eq:Wmu-tau}) and~(\ref{eq:Wmu-tau'st}), respectively,
after the substitution $\Delta m_2^2 \leftrightarrow \Delta m_1^2$.
The total probability of the $\nu_e$ oscillation transformations reads
\begin{equation}
W(\nu_e \to \nu_s + \nu_\tau + \nu_\mu) =
\cos^4 \vartheta \sin^2 (\delta m_1^2) +
\sin^4 \vartheta \sin^2 (\delta m_2^2) +
\sin^2 2\vartheta < \sin^2 (\delta m_{12}^2) >.
\label{eq:We-tau'st'mu}
\end{equation}
The first term in this equation is the dominant one with nearly maximal
amplitude. It comes from the transformations~$\nu_e \to \nu_s$
and~$\nu_e \to \nu_\tau$, their partial contributions are with
factors~$\cos^2 \varphi$ and~$\sin^2 \varphi$, respectively.
Again, the contribution from the transformations $\nu_e \to \nu_\mu$
is a small short-baseline effect.

The phenomenology above leads to a complementary relation between the
possible dominant modes in the atmospheric and the solar oscillations:
if the tau neutrino dominates the atmospheric $\nu_\mu$-oscillations,
then the sterile neutrino should dominate the solar $\nu_e$-oscillations,
and conversely.

If the second mixing angle in the ansatz~(\ref{eq:nu_e})--(\ref{eq:nu_s})
is very small, $\varphi \ll 45^\circ$, as in the economical one-parameter
version~(\ref{eq:neutrinos}), the tau neutrino will strongly
dominate the atmospheric $\nu_\mu$-oscillations, and the sterile
neutrino will strongly dominate the solar $\nu_e$-oscillations. This
inference is in good agreement with the recent atmospheric Super-Kamiokande
muon neutrino oscillation data~\cite{Fukuda}, but it does not agree with
the solar neutrino data~\cite{Bahcall,Suzuki}, which seem to exclude the
$\nu_e \to \nu_s$ oscillation mode as the dominant one in the solar
neutrino oscillations.

In the other extreme case with a large mixing angle $\varphi = 45^\circ$,
the dominant disappearance neutrino transformations in both the atmospheric
and solar neutrino oscillations should be the same,
$\nu_\mu \to (\nu_\tau + \nu_s)$ and $\nu_e \to (\nu_\tau + \nu_s)$,
respectively, with equal~$\nu_\tau$ and~$\nu_s$ modes in each case,
while the $\nu_\mu \leftrightarrow \nu_e$ modes are strongly suppressed
in these long-baseline neutrino oscillations in accordance at least with
the most confident atmospheric Super-Kamiokande data~\cite{Super-Kamiokande}.

Note, that none of the solar neutrino oscillation data are comparable
with the Super-Kamiokande atmospheric neutrino oscillation
data~\cite{Super-Kamiokande} by their confidence as
yet~\cite{Bahcall,Suzuki}. With the new atmospheric neutrino
indication~\cite{Fukuda}, the mixing angle region $\varphi \lesssim 45^\circ$
is certainly the least disfavored by the recent neutrino oscillation data.

\section{Conclusion}
\label{sec:conclusion}

To conclude, an attempt is made to explain relations between well known
different neutrino data by the suggestion of a new unifying physical
notion, the Dirac-Majorana neutrino duality. The nearly maximal neutrino
transformation $\nu_\mu \to (\nu_\tau + \nu_x)$, $x \neq e$,
oscillation amplitude in the atmospheric Super-Kamiokande neutrino
experiment~\cite{Super-Kamiokande}, together with lepton charge
conservation in all the known weak interaction reactions, and the data
dictated doublet character of the neutrino mass spectrum,
Eq.~(\ref{eq:mass-differences}), with the LSND indications accepted,
e.g.~\cite{Bilenky-99}, prompts an extended analogy between the neutrino
properties and the duality properties of the neutral kaons. Duality means
an indication of new physics: the neutrinos \textit{are} carrying lepton
charge Dirac fields in the semiweak interactions, and they \textit{are}
pairs of dual truly neutral Majorana fields in the long-baseline neutrino
oscillations. The massive neutrinos are produced in the semiweak
interactions as Dirac particles with not complete longitudinal polarization,
and they remain nearly Dirac neutrinos (see the discussion of
Eq.~(\ref{eq:nu-eigenstate1}$^\prime$)) at finite distances from the
production vertex until these distances remain much shorter than the
long-baseline oscillation lengths.

The necessary restriction to three active in the weak interactions
lepton flavors determines a minimum of two four-component massive Dirac
neutrinos, two associated pairs of Majorana neutrinos, and it leads to
the involvement of one sterile neutrino which has to be imparted in the
neutrino oscillations.

The suggested dual Dirac-Majorana properties of the neutrinos are
modelled in a minimal four-neutrino mixing ansatz for the three flavor
and one sterile neutrinos~(\ref{eq:nu_e})--(\ref{eq:nu_s}) and
the Dirac neutrino mass term~(\ref{eq:mass-termD}) which describe lepton
charge conservation in the semiweak interactions with a special lepton
charge pattern~(\ref{eq:leptons-7}), plus a Majorana neutrino mass
term~(\ref{eq:mass-termM}) which is regarded by definition as a small
perturbation effect. As a result, we get a simple unifying neutrino
oscillation phenomenology with two nearly mass-degenerate Majorana
neutrino doublets in Eq.~(\ref{eq:Lagr-total}), the maximal neutrino mixing
condition in Eq.~(\ref{eq:substitutions}) and the physically motivated
important restriction to not more than two mixing angles~$\vartheta$
and~$\varphi$ in the CP-invariant lepton semiweak interactions
with a symmetric $(2 + 2)$ neutrino mixing
pattern~(\ref{eq:nu_e})--(\ref{eq:nu_s}) plus~(\ref{eq:substitutions}).
The fine neutrino mass doublet structure is natural here because it is
produced by a lepton charge symmetry violating perturbation on a
zero-approximation system of two exactly twofold mass-degenerate Dirac
neutrino-antineutrino pairs.

The model implies violation of the $(e, \mu)-\tau$ universality and allows
a group of suppressed by the small Dirac neutrino masses reactions,
which involve basically the helicity-flip lepton charge conserving
Dirac neutrino transformations
$\nu_\mu (\nu_e) \leftrightarrow \tilde \nu_\tau$
(in contrast to the lepton charge violating, $\Delta l = \pm 2$,
helicity conserving Majorana neutrino transformations
$\nu_\mu (\nu_e) \leftrightarrow \nu_\tau$
in the long-baseline neutrino oscillations), which will be testable
in the next generations of neutrino reaction experiments.
If the effective Dirac mass of the muon neutrino is not in the eV, but in
the 100~keV region indeed~\cite{PDG}, the production of the
antitauons~$\tau^+$ by the high energy muon neutrinos~$\nu_\mu$ on nuclei
is more interesting, it is singled out in Sec.~\ref{sec:DM-ansatz}.

The necessary involvements of the sterile neutrino in the solar and
atmospheric neutrino oscillations will be tested in many coming neutrino
oscillation experiments as SNO, Super-Kamiokande, MiniBooNE, KARMEN, the
accelerator long-baseline K2K and MINOS experiments and other
neutrino oscillation experiments~\cite{PANIC-99}.

A crucial test of the ansatz~(\ref{eq:nu_e})--(\ref{eq:nu_s})
will be the coincidence of the two seemingly independent experimental
values of the mixing angle~$\varphi$ from the long-baseline accelerator
neutrino oscillation transformations $\nu_\mu \to \nu_\tau$,
Eq.~(\ref{eq:Wmu-mu}), and the short-baseline $\nu_\tau$-survival
oscillations, Eq.~(\ref{eq:Wtau-st}). With Dirac-Majorana neutrino duality,
the restriction to not more than two mixing angles is feasible in the
phenomenology of neutrino oscillations at the approximation of CP-invariance.

Beyond the formal differences, the suggested duality properties of the
neutrinos are analogous to the well known duality properties of the neutral
kaons. This physical idea was initiated by Pontecorvo and has gained new
support~\cite{Lipmanov-98,Lipmanov-99} from the leading neutrino oscillation
data since the main Super-Kamiokande atmospheric neutrino oscillation
results in 1998~\cite{Super-Kamiokande}, and also the LSND
indications~\cite{Athanassopoulos}. The new versatile developments in
experimental neutrino physics will eventually prove or disprove
this physical idea finally, e.g.~\cite{(2+2)-mixing scheme}.

\section*{Acknowledgements}

I would like to thank N.V.~Mikheev for very useful discussions and valuable
critical comments and L.A.~Vassilevskaya and A.Ya.~Parkhomenko for interest
in this work and preparation of the LaTeX file.

\end{document}